\documentclass[12pt]{article}
 \usepackage{amsmath,amsfonts,latexsym,amssymb,amsthm}

   \newcommand{\field}[1]{\mathbb{#1}}
   \newcommand{\rz}{\field{R}}
   \newcommand{\cz}{\field{C}}

   \newcommand{\dd}[1]{\frac{\partial}{\partial #1}}
   
   \newcommand{\ab}[2]{\frac{\partial #1}{\partial #2}}
   \newcommand{\dirac}{/\!\!\!\nabla}
   
   \newcommand{\daggern}{/\!\!\!n}

 \newtheorem{theorem}{Theorem}[section]
 \newtheorem{definition}[theorem]{Definition}
 
 \newtheorem{lem}[theorem]{Lemma}
 
 \newtheorem{pro}[theorem]{Proposition}

\begin{document}

\title{The Reeh-Schlieder Property for the Dirac Field on Static Spacetimes}

\author{Alexander Strohmaier}

\date{\small Universit\"at Leipzig,
Institut f\"ur theoretische Physik,
Augustusplatz 10/11, D-04109 Leipzig, Germany\\
E-mail: alexander.strohmaier@itp.uni-leipzig.de}

\maketitle

\begin{abstract}
  We prove the Reeh-Schlieder property for the ground- and KMS-states states
  of the massive Dirac Quantum field on a static globally hyperbolic 4 dimensional
  spacetime.
\end{abstract}

\maketitle

\newpage

\section{Notations}

We use as a standard representation of the Clifford algebra the Dirac matrices
$$
\gamma^0=
\left(
\begin{array}{cc} \mathbf{1}_2 & 0 \\
                        0     & -\mathbf{1}_2
\end{array} \right)
$$
$$
\gamma^i=\left(
\begin{array}{cc} 0 & \sigma_i \\
                -\sigma_i     & 0
\end{array} \right)
$$
$$
\quad i=1,2,3
$$
where $\sigma_i$ are the Pauli matrices.\vspace{0.5cm} \\
If $M$ is a smooth manifold and $E$ a smooth vector bundle over $M$ we use the
following notations\\
$ \Gamma(E) \quad \ldots$ space of smooth sections\\
$ \Gamma_0(E) \quad \ldots$ space of compactly supported smooth
  sections\vspace{0.5cm}\\
If $M$ is a time-oriented oriented Lorentz manifold and $\mathcal{O}$ a
subset of $M$ we use\\
$J^\pm(\mathcal{O}) \quad \ldots$ set of points which can be reached
  by future/past directed causal curves emanating from  $\mathcal{O}$.\\
$J(\mathcal{O})=J^+(\mathcal{O}) \cup J^-(\mathcal{O})$. \\
$D^\pm(\mathcal{O}) \quad \ldots$ set of points $p \in J^\pm(\mathcal{O})$
such that every past/future inextendible causal curve starting at $p$ passes
through $\mathcal{O}$.\\
$D(\mathcal{O})=D^+(\mathcal{O}) \cup D^-(\mathcal{O})$.\\
Let $\mathcal{O}$ be an open subset of $M$:\\
$\mathcal{O}^\perp = Int(D(\mathcal{O})^c) \quad \ldots$ causal complement of
$\mathcal{O}$.\vspace{0.5cm}\\
If $\mathcal{H}$ is a Hilbert space and $\mathcal{E} \subset
\mathcal{L}(\mathcal{H})$:\\
$ \mathcal{E}' \quad \ldots$ commutant of $\mathcal{E}$, i.e. 
$ \mathcal{E}':=\lbrace a \in \mathcal{L}(\mathcal{H}); [x,a]=0 \quad \forall x \in
\mathcal{E}\rbrace$.

\newpage

\section{Introduction}

For the analysis of Quantum Field Theory (QFT) in curved spacetime it has
turned out that the framework of algebraic Quantum Field Theory (AQFT) is most suitable for analyzing
the problems connected with the non-uniqueness of the quantization of linear
fields (see e.g. \cite{Wald} and the references therein).
The problem reduces to finding appropriate representations of the field
algebra, which can be constructed straightforwardly on manifolds (see \cite{Dimock:1980hf,Dimock:1982hf}).
This is automatically obtained by specifying a vacuum state over the algebra.
On static spacetimes it is possible to distinguish such vacuum states
as ground states with respect to the canonical time translations.

In the Minkowski space theory the vacuum turns out to be cyclic for field
algebras associated to non-void open regions (\cite{Reeh:1961re}). This Reeh-Schlieder property
of the vacuum provides the starting point for the use of Tomitas and
Takesakis Modular theory in AQFT. By the lack of symmetry in general
spacetimes it is not clear whether physically reasonable vacuum states have
this property as well. It was shown in \cite{Verch:1993pn} that the quasifree ground
state of the massive scalar field on an ultrastatic spacetime is of this
kind.
The proof makes use of an anti-locality property of the square root
of the Laplace operator. We show here, using a similar method, that in the
case of the free massive Dirac field on a static spacetime any
KMS or ground state with respect to the
canonical time translation has the Reeh-Schlieder property.

\section{The free Dirac field on a manifold}

Let $M$ be a 4 dimensional connected oriented time-oriented Lorentz manifold which
admits a smooth Cauchy surface. It is known that $M$ possesses a trivial spin
structure, given by a $\textrm{Spin}^+(3,1)$-principal bundle $SM$ and a two fold covering
map $SM \to FM$ onto the bundle $FM$ of oriented time-oriented frames.
One can construct now the Dirac bundle $DM$, which is associated to
$SM$ by the spinor representation and is a natural module for the Clifford
algebra bundle Cliff($TM$) (see \cite{Baum} and \cite{Lawson}).
Furthermore the Levi-Civita connection on $TM$
induces a connection on $DM$ with covariant derivative
$\nabla : \Gamma(SM) \to \Gamma(SM \otimes T^*M)$.
Given a vector field $n$ we write as usual $\daggern$ for the section in the
Clifford algebra bundle $\gamma(n)$, or in local coordinates $\gamma^i n_i$. 
We denote by $D^*M$ the dual bundle of $DM$.
There exists an antilinear bijection
$\Gamma(DM) \to \Gamma(D^*M): u \to u^+$, the Dirac conjugation, which
in the standard representation in a local orthonormal spin frame has the form
$u^+=\overline{u} \gamma^0$, where the bar denotes complex conjugation in
the dual frame.
We use the symbol ${}^+$ also for the inverse map.
Canonically associated with the Dirac bundle there is the Dirac operator
which in a frame takes the form
$$ \dirac = \gamma^i \nabla_{e_i}\;.$$
The Dirac equation for mass $m >0 $ is
\begin{equation}
 (-i \dirac + m) u = 0, \quad u \in \Gamma(DM)\;.
\end{equation}

The proofs of the following facts can be found in \cite{Dimock:1982hf}.
We state these results here as propositions.

\begin{pro}\label{cauchy}
 Under the stated assumptions $ -i \dirac + m $ on $\Gamma(DM)$ has unique
 fundamental solutions $S^\pm: \Gamma_0(DM) \to \Gamma(DM)$, satisfying
 $$ (-i \dirac + m) S^\pm = S^\pm (-i \dirac + m) = id \quad \textrm{on}
 \quad \Gamma_0(DM)\;,$$
 $$ \textrm{supp}(S^\pm f) \subset J^\pm(\textrm{supp}(f))\;.$$
\end{pro}

We can define the operator $S:=S^+-S^-$.
There is also a positive answer to the Cauchy problem.
Let $\Sigma$ be a smooth Cauchy surface and let $D\Sigma$ the restriction of
the bundle $SM$ to $\Sigma$ with restriction map $\rho_\Sigma$.

\begin{pro}
 For each $u_0 \in \Gamma_0(D\Sigma)$ there exists a unique $u \in
 \Gamma(DM)$ such that
 $$ (-i \dirac + m) u = 0, \quad \rho_\Sigma(u)=u_0\;.$$
 Furthermore $\textrm{supp}(u) \subset J(\textrm{supp}(u_0))$.  
\end{pro}

We form the prehilbert space $\mathcal{K}:=\Gamma_0(DM)/\textrm{ker}(S)$
with inner product
\begin{gather}\label{fip}
\langle [u_1], [u_2] \rangle_\mathcal{K} := -i \int_M (u_1^+, S u_2)(x) d\mu(x),
\end{gather}
where $(\cdot,\cdot)$ denotes the dual pairing between the fibres of $DM$
and $D^*M$ and $\mu(x)$ the canonical (pseudo)-riemannian measure on $M$.
$[f]$ denotes the equivalence class containing $f$.
Analogously we form the prehilbert space
$\mathcal{K^+}:=\Gamma_0(D^*M)/(\textrm{ker}(S))^+$ with inner product
\begin{gather}\label{sip}
  \langle [v_1], [v_2] \rangle_\mathcal{K^+} := \langle [v_2^+], [v_1^+] \rangle_\mathcal{K}\;.
\end{gather}
We note that $S$ maps $\mathcal{K}$
onto the space of smooth solutions of the Dirac equation whose support have compact intersection
with any Cauchy surface.
The construction of the Dirac field starts with the Hilbert space
$\mathcal{H}:= \overline{\mathcal{K} \oplus \mathcal{K}^+}$ and
the antiunitary involution
\begin{gather}
  \Gamma: \mathcal{H} \to \mathcal{H}, \quad u \oplus v \to v^+\oplus u^+\;.
\end{gather}
The field algebra $\mathcal{F}$ is the (self-dual) CAR-algebra
$\textrm{CAR}(\mathcal{H}, \Gamma)$. 
This is the $C^*$-algebra with unit generated by symbols $B(g)$ with
$g \in\mathcal{H}$ and relations
\begin{gather}
g \to B(g) \quad \textrm{is complex linear,}\\
B(g)^* = B(\Gamma g),\\
\lbrace B(g_1) , B(g_2) \rbrace = B(g_1)B(g_2)+B(g_2)B(g_1)=\langle \Gamma g_1,g_2 \rangle\;.
\end{gather}
For each relatively compact subset $\mathcal{O} \subset M$ we define the
local algebra $\mathcal{F}(\mathcal{O}) \subset \mathcal{F}$ to be the
closed unital $*$-subalgebra generated by the symbols $B([f])$ with $f \in
\Gamma_0(DM) \oplus \Gamma_0(D^*M)$ and $\textrm{supp}(f) \subset
\mathcal{O}$.
It was shown in \cite{Dimock:1982hf} that the assignment
$$ \mathcal{O} \to \mathcal{F}(\mathcal{O})$$
defines a local, causal, covariant net of field algebras over $M$.

\section{Stationary Spacetimes}

In addition to the assumptions of the previous section we assume now that
$M$ is static, i.e. there exists a one-parameter
group $\alpha_t$ of isometries of $M$ with timelike orbits and a spacelike
hypersurface which is orthogonal to the orbits.
The corresponding complete timelike Killing vector field will be denoted by
$\xi$. The push-forward $(\alpha_t)_*: TM \to TM$ induces an equivariant flow
on the $\mathcal{L}_0^\uparrow$-principal bundle $FM$.
Since $SM$ is a double cover of $FM$, the flow on $FM$ lifts
uniquely to a flow $\tilde\beta_t$ of the $\textrm{SL}_2(\cz)$-principal bundle $SM$,
which is again equivariant and hence induces a flow $\beta_t$ on the
Dirac-bundle $DM$. $\beta_t$ gives rise to a linear transformation group
of $\Gamma(DM)$, which commutes with the Dirac operator (see \cite{Dimock:1982hf}). Similarly
${}^+ \circ \beta_t \circ {}^+$ defines a linear transformation group of
$\Gamma(DM^*)$.
Both groups leave $\textrm{ker}(S)$ and $\textrm{ker}(S)^+$
invariant and define linear transformation groups of $\mathcal{K}$ and
$\mathcal{K}^+$, which preserve the scalar products (\ref{fip}) and
(\ref{sip}) (see \cite{Dimock:1982hf}).
This finally gives a one-parameter group of unitary operators $U(t):
\mathcal{H} \to \mathcal{H}$.

\begin{pro}
 The group $U(t)$ is strongly continuous and commutes with the conjugation
 $\Gamma$. It therefore defines a strongly continuous group $\tau_t$ of
 Bogoliubov automorphisms of $\mathcal{F}$. Moreover we have covariance with
 respect to these group actions, i.e. 
 $\tau_t\mathcal{F}(\mathcal{O})=\mathcal{F}(\alpha_t \mathcal{O})$
\end{pro}
\begin{proof}
For one parameter groups weak continuity implies strong continuity.
It is clearly sufficient to show that the group is weakly continuous on
$\overline{\mathcal{K}}$, viewed as a subspace of $\mathcal{H}$, which is
equivalent to
\begin{gather}
  \lim_{t \to 0} \langle U(t)v-v,  u \rangle_\mathcal{K} = 0 \quad \forall u,v \in \mathcal{K}\;. 
\end{gather}
Given $f,g \in \Gamma_0(DM)$ we have
\begin{gather}
 \lim_{t \to 0}\langle U(t)[f]-[f], [g] \rangle_\mathcal{K} =-i \lim_{t \to 0}
 \int_M h(x,t) d\mu(x),
\end{gather}
where
\begin{gather}
 h(x,t) := ((\beta_t f-f)^+,S g)(x).
\end{gather}
We fix a finite interval $I=[-\epsilon,\epsilon]$ and find that
$\mathcal{V}:=\cup_{t \in I} \alpha_t(\textrm{supp}(f))$ is compact.
For each $t \in I$ we therefore have $\textrm{supp}(h(\cdot,t)) \subset
\mathcal{V}$. This implies that $h$ is bounded as a function on $M \times I$,
and therefore there exists a function
$\tilde h \in L^1(M)$, such that $\vert h(x,t) \vert \leq
\vert \tilde h(x)\vert \quad \forall t \in I$.
Hence, we can perform the limit under the integral, and
since $\lim_{t \to 0} h(x,t) = 0$, weak continuity follows.
The group $U(t)$ commutes by construction with $\Gamma$.
It is now not difficult to check on the generators that the corresponding
group of Bogoliubov automorphisms is strongly continuous. The covariance is
clear by construction.
\end{proof}

We interpret the automorphism group $\tau_t$ as the group of
time-translations canonically given on a static spacetime, and it seems
physically reasonable to investigate states which satisfy certain stability
requirements with respect to this automorphism group.
We are thus interested in the following class of states.

\begin{definition}
Let $\mathcal{A}$ be a $C^*$-algebra with unit and $\tau_t$
be a strongly continuous one parameter group of $*$-automorphisms of $\mathcal{A}$.
A $\tau_t$ invariant state $\omega$ is called ground state
with respect to $\tau_t$, if the generator of the corresponding
unitary group $\tilde U(t)$ on the GNS Hilbert space is positive.
\end{definition}
\begin{definition}
Let $\mathcal{A}$ and $\tau_t$ be as above.
A state $\omega$ is called KMS state with inverse temperature $\beta>0$
with respect to $\tau_t$, if for any pair $A,B \in \mathcal{A}$
there exists a complex function $F_{A,B}$ which is analytic in the strip
$$\mathcal{D}_\beta:=\lbrace z \in \cz; 0<Im(z)<\beta \rbrace$$ and bounded
and continuous on $\overline{\mathcal{D}_\beta}$, such that
\begin{gather*}
 F_{A,B}(t)=\omega(A \tau_t(B))\\
 F_{A,B}(t+i\beta)=\omega(\tau_t(B)A)\;.
\end{gather*}
\end{definition}

\section{The Reeh-Schlieder property for ground and KMS states}

\begin{definition}
 Let $\lbrace \mathcal{F}(\mathcal{O}) \rbrace$ be a net of local
 field algebras indexed by the relatively compact open sets of a manifold $M$.
 Let $\omega$ be a state over the field algebra $\mathcal{F}$,
 and $(\pi_\omega, \mathcal{H}_\omega,\Omega_\omega)$ its GNS triple.
 We say that $\omega$ has the Reeh-Schlieder property, if $\Omega_\omega$
 is cyclic for the von Neumann algebra $\pi_\omega(\mathcal{F}(\mathcal{O}))''$ for
 any non-void open relatively compact set $\mathcal{O}$.
\end{definition}

We show in this section the following theorem:
\begin{theorem}\label{rsp}
 Let $M$ be a connected orientable time-orientable static globally
 hyperbolic 4 dimensional Lorentzian manifold.
 Let $\lbrace \mathcal{F}(\mathcal{O}) \rbrace$ be the net of local field
 algebras for the free Dirac field with mass $m>0$ and let $\omega$ be
 a ground or KMS state with respect to the automorphism group $\tau_t$,
 induced by the timelike Killing flow.
 Then $\omega$ has the Reeh-Schlieder property.
\end{theorem}

Let $\mathcal{O} \subset M$ be a nonvoid open relatively compact set.
We denote by $\mathcal{B}(\mathcal{O}) \subset \mathcal{F}(\mathcal{O})$ the
$*$-subalgebra consisting of those elements $a \in \mathcal{F}(\mathcal{O})$ for which there exists a
neighbourhood $I \subset \rz$ of $0$, such that $\tau_I(a)
\subset \mathcal{F}(\mathcal{O})$. Let $\mathcal{O}' \subset M$ be another nonvoid
open set, such that $\overline{\mathcal{O}'} \subset \mathcal{O}$. We clearly
have the inclusions
\begin{gather}\label{incl}
 \mathcal{F}(\mathcal{O}') \subset \mathcal{B}(\mathcal{O})
 \subset \mathcal{F}(\mathcal{O})\;.
\end{gather}
We denote by $\tilde U(t)$ the strongly continuous unitary
group, implementing $\tau_t$ on the GNS-Hilbert space.
One has
\begin{lem}\label{invG}
 Let $\mathcal{E}:=\overline{\pi_\omega(\mathcal{B}(\mathcal{O}))
 \Omega_\omega}=\overline{\pi_\omega(\mathcal{B}(\mathcal{O}))'' \Omega_\omega}$. 
 Then $\mathcal{E}$ is invariant under the action of $\tilde U(t)$, i.e.
 $\tilde U(\rz) \mathcal{E} \subset \mathcal{E}$.
\end{lem}  
\begin{proof}
Let $\psi \in \mathcal{E}^\perp$. Then for each $a \in
\pi_\omega(\mathcal{B}(\mathcal{O}))$ we
have at least for some open neighbourhood $I \subset \rz$ of 0:
$$
 f(t):=\langle \psi, \tilde U(t) a\Omega_\omega \rangle=0 \quad \forall t \in I\;.
$$
Since $\omega$ is a KMS-state ($\beta>0$) or a ground state ($\beta=\infty$),
$f(t)$ is the boundary value of a function $F(z)$, which is analytic
on the strip $$\mathcal{D}_{\beta/2}=\lbrace z \in \cz; 0<Im(z)<\beta/2
\rbrace$$ and bounded and continuous on $\overline{\mathcal{D}_{\beta/2}}$.
By the Schwartz reflection principle $f(t)$ vanishes on the whole real axis.
Therefore $\langle \tilde U(t) \psi, a\Omega_\omega \rangle=0$ for all
$t \in \rz$. 
Hence, $\mathcal{E}^\perp$ is invariant under the action of $\tilde U(t)$ and the lemma follows.
\end{proof}

\begin{lem}
 Let $\mathcal{H}'$ be the subspace of $\mathcal{H}$ generated by the set
 $$
 \lbrace [f] \in \mathcal{H}; f \in \Gamma_0(DM) \oplus
 \Gamma_0(D^*M),\textrm{supp}(f) \subset \alpha_\rz \mathcal{O}' \rbrace,
 $$
 and let
 $$
 \mathcal{R}:=\lbrace \pi_\omega(B(f)), f \in \mathcal{H}' \rbrace'' \subset
 \pi_\omega(\mathcal{F})''\;.
 $$
 Then $\mathcal{E}$ is invariant under the action of $\mathcal{R}$, i.e. $\mathcal{R}\mathcal{E}
 \subset \mathcal{E}$.
\end{lem}
\begin{proof}
By the inclusion (\ref{incl}) and Lemma \ref{invG} $\mathcal{E}$ is invariant
under the action of $\bigvee_{t \in \rz}
\pi_\omega\left(\tau_t(\mathcal{F}(\mathcal{O}'))\right)''$.
Since the set $\lbrace B([f]); \textrm{supp}(f) \subset \mathcal{O}' \rbrace$ generates the algebra $\mathcal{F}(\mathcal{O}')$ we
have $\mathcal{R}=\bigvee_{t \in \rz}
\pi_\omega\left(\tau_t(\mathcal{F}(\mathcal{O}'))\right)''$.
\end{proof}
Our goal is now to show that
 $\mathcal{H}'$ is dense in $\mathcal{H}$.\\
$(\mathcal{H}')^\perp$ is clearly $U(t)$-invariant and hence
contains a dense set of analytic vectors for $U(t)$.
Let $\phi$ be such an analytic vector.
We take the two components of $\phi_{1,2}$ of $\phi$ such that $\phi_1 \in
\overline{\mathcal{K}}$ and  $\phi_2 \in \overline{\mathcal{K}^+}$.
We will show now that $\phi_1=0$, and the proof of $\phi_2=0$ will be analogous.
Thinking of $\overline{\mathcal{K}}$ as a subspace of $\mathcal{H}$, we clearly have
\begin{gather}\label{support}
 \langle U(it) \phi_1 , [f] \rangle =0 \quad \forall t \in \rz, f \in
 \Gamma_0(DM) \textrm{ with } \textrm{supp}(f) \subset \mathcal{O}'\;,
\end{gather}
The bundle $\Gamma_0(DM)$ is trivial, and for each trivialization the map
\begin{gather}
  \Gamma_0(DM) \to \cz, \quad f \to \langle \phi_1,[f]
  \rangle
\end{gather}
defines a distribution $\tilde\psi \in \mathcal{D}'(M) \otimes \cz^4$.
Let $\mathcal{V}$ be the closure of the open set $M \setminus \textrm{supp}(\tilde\psi)$. 
By equation (\ref{support}) it is clear that $\mathcal{O}' \subset
\mathcal{V}$. Hence, $\textrm{Int}(\mathcal{V})$ is nonvoid.
We will show that $\mathcal{V}$ is open, and therefore $\mathcal{V}=M$.

Since the Killing vector field $\xi$ is nowhere vanishing, for each point
$p \in \mathcal{V}$ there is an open neighbourhood $\mathcal{U}$ of $p$ and a chart
$\mathcal{U} \to \rz^4$
with coordinates $(x_0,x_1,x_2,x_3)$, such that $\xi=\dd{x_0}$.
As $M$ is static we can choose the chart such that the coefficients of the metric tensor
are independent of $x_0$, and $g_{0i}=g_{i0}=0$ for
$i \in \lbrace 1,2,3 \rbrace$.
Without loss of generality we can assume that the chart maps $\mathcal{U}$
to a region of the form $I \times \mathcal{U}'$, where
$I=(-\epsilon,\epsilon) \subset \rz$
is an open interval, and $\mathcal{U}'$ is a bounded open connected subset of $\rz^3$.
The chart locally trivialises the tangent bundle, and we can choose an
orthonormal time-oriented oriented frame $t_k$, such that the Lie-Derivatives
$\mathcal{L}_\xi(t_k)$ vanish. It is not difficult to see that this implies
the existence of a local orthonormal spin frame $e_i$, with
$\mathcal{L}_\xi(e_i):=\lim_{t \to 0} \frac{\beta_{-t} e_i -e_i}{t}=0$.
We use this frame to identify the sections of $\Gamma_0(DM)$  which are
supported in $\mathcal{U}$ with $C_0^\infty(I \times \mathcal{U}') \otimes \cz^4$.
If $f \in C_0^\infty(I/2 \times \mathcal{U}') \otimes \cz^4$
we have in our frame $\beta_t f(x_0,x_i) = f(x_0-t,x_i)$
for all $t \in I/2$.

\begin{lem}\label{vanish}
 $\mathcal{V}':=(I \times \mathcal{U}') \cap \textrm{Int}(\mathcal{V})$ contains an
 open set of the form $I \times \tilde\mathcal{U}$, where $\tilde\mathcal{U}
 \subset \mathcal{U}'$ is open and nonvoid.
\end{lem}
\begin{proof}
$\mathcal{V}'$ is open and nonvoid. Let $\mathcal{V}'' \subset
\mathcal{V}'$ be another nonvoid open set, such that
$\alpha_{(-\delta,\delta)}\mathcal{V}'' \subset \mathcal{V}'$ for some $\delta
> 0$. The function $F_f(t):=\langle U(t) \phi_1, [f] \rangle$
is entire analytic in $t$ and vanishes on $(-\delta,\delta)$ if $f \in
\Gamma_0(DM))$ with $\textrm{supp}(f) \subset \mathcal{V}''$.
Therefore we get $F_f=0$ whenever
$\textrm{supp}(f) \subset \mathcal{V}''$, and as a consequence
$\alpha_\rz \mathcal{V}'' \subset \textrm{Int}(\mathcal{V})$.
This already proves the lemma, since in our chart $\alpha_t$ is just the
shift $x_0 \to x_0+t$.
\end{proof}
$\phi_1$ defines via the maps
\begin{gather}
  \left( C_0^\infty(I/2 \times \mathcal{U}') \otimes \cz^4 \right) \otimes
  C_0^\infty(I/2) \to \cz, \nonumber\\
  \quad f \otimes g \to \int_\rz \langle U(it) \phi_1, [f] \rangle \; g(t)
  dt\\
  \left( C_0^\infty(I/2 \times \mathcal{U}') \otimes \cz^4 \right) \otimes
  C_0^\infty(I/2) \to \cz, \nonumber\\
  \quad f \otimes g \to \int_\rz \langle U(t) \phi_1, [f] \rangle \; g(t) dt
\end{gather}
distributions $\psi, \hat\psi \in \mathcal{D}'(I/2 \times \mathcal{U}' \times
I/2) \otimes \cz^4$.
For an arbitrary $g \in C_0^\infty(I/2)$ we define the distributions
$\psi_g,\hat\psi_g \in \mathcal{D}'(I/2 \times \mathcal{U}') \otimes \cz^4$ by smearing in the first
variable with g:
\begin{gather}
  \psi_g:=\sigma ( \psi(g,\cdot,\cdot) )\\
  \hat\psi_g:=\sigma ( \hat\psi(g,\cdot,\cdot) ),
\end{gather}
where $\sigma$ is the flip identifying
$\mathcal{D}'(\mathcal{U}' \times I/2) \otimes \cz^4$ with $\mathcal{D}'(I/2
\times \mathcal{U}') \otimes \cz^4$.

\begin{lem}
  $\psi_g=0$.
\end{lem}
\begin{proof}
The distribution $\langle \phi_1 , [\cdot]\rangle \in \mathcal{D}'(I/2 \times
\mathcal{U}') \otimes \cz^4$ is a distributional solution to the Dirac equation,
and by Lichnerowics formula a solution to the equation
\begin{gather}
 (\square_\mathcal{S}+m^2-1/4 R) h = 0,
\end{gather}
where $\square_\mathcal{S}$ is the spinor Laplace and $R$ the scalar
curvature. The map $U(t)$ shifts in $x_0$ direction, and it is easy to see
that therefore $\hat\psi_g$ solves this equation as well.
We note that all coefficients of the differential equation are independent of
$x_0$, and the principal part of the operator $\square_\mathcal{S}$
is diagonal with real coefficients.
If we replace in $\square_\mathcal{S}$ all derivatives $\dd{x_0}$
by $-i\dd{x_0}$ we obtain an elliptic second order differential operator
$\square_\mathcal{S}^{\textrm{eukl}}$
and clearly $\psi_g$ solves the equation
\begin{gather}
  (\square_\mathcal{S}^{\textrm{eukl}}+m^2-1/4 R) h = 0.
\end{gather}
Hence, $\psi_g$ is smooth (see e.g.\cite{HormBook:1990,Dencker:1981de}).
If $v_i$ are the components of $\psi_g$ we can therefore
write the equation
in the form
\begin{gather}
  A v_i = \sum_{j=1}^4 B_{ij} v_j \quad i=1,2,3,4\;,
\end{gather}
such that A is an elliptic second order differential operator, and $B_{ij}$
is a Matrix of first order differential operators.
It follows that the $v_i$ satisfy the inequalities
\begin{gather}
 \vert A v_i \vert^2 \leq N \sum_{l=1}^4 \left( \sum_{k=0}^3 \vert
 \ab{v_l}{x^k} \vert^2 + \vert v_l \vert^2 \right)
\end{gather}
on $I/2 \times \mathcal{U}'$ for some constant $N \geq 0$.
Furthermore, by lemma \ref{vanish}, the $v_i$ vanish on the open nonvoid set
$I/2 \times \tilde\mathcal{U}$.
Hence, by the result of \cite{Aroz:1957ar} (see especially \mbox{Remark 3})\footnote{
For a detailed treatment see section 17.1 of \cite{Hormbook:1985} and the
references therein} we get $v_i=0$
and $\psi_g=0$.
\end{proof}

Since $\psi_g$ vanishes for all $g \in C_0^\infty(I/2)$ it follows
that $\psi=0$ and therefore $I/2 \times \mathcal{U}' \subset
\textrm{Int}(\mathcal{V})$. As a consequence $\mathcal{V}$ is open
and $\mathcal{V}=M$. This completes the proof that $\phi_1=0$.
With the same arguments we can show $\phi_2=0$ and hence $\phi=0$.
It follows that $(\mathcal{H}')^\perp = \lbrace 0 \rbrace$ and we have shown
\begin{lem}
 $\mathcal{H}'$ is dense in $\mathcal{H}$.
\end{lem}
\hspace{0.1cm}\\
\it Proof of theorem \ref{rsp}. \rm
Since $\mathcal{H}'$ is dense in $\mathcal{H}$ we get immediately
$\mathcal{R}=\pi_\omega(\mathcal{F})''$. Since $\Omega_\omega$ is cyclic for
$\pi_\omega(\mathcal{F})''$ we have $\mathcal{E}=\mathcal{H}_\omega$ and with inclusion
(\ref{incl})
we get
\begin{gather*}
\mathcal{E} \subset \overline{\pi_\omega(\mathcal{F}(\mathcal{O}_\Sigma))''
  \Omega_\omega}=\mathcal{H}_\omega\;.
\end{gather*}
\endproof

\section{Conclusions and Outlook}

The Reeh-Schlieder property may serve for further investigation
of the Dirac field on static spacetimes.
Candidates for vacuum states on static spacetimes like
the Rindler spacetime and the external space of the black-hole solution
are typically KMS-states, and our theorem applies to these cases.
It would also be desireable to establish this property
for classes of vacuum states on algebras associated
to fields with higher spin or massless fields.
Since our proof makes use of certain properties of the CAR algebra
it is not clear to us yet whether an analogous statement holds
for the bosonic case.

\section{Acknowledgements}
The author would like to thank Prof. M. Wollenberg for useful discussions.
This work was supported by the
Deutsche Forschungsgemeinschaft within the scope of the postgraduate
scholarship programme ``Graduiertenkolleg Quantenfeldtheorie'' at the
University of Leipzig.

\end{document}